\documentclass{article}

\usepackage[preprint]{main}
\makeatletter
\renewcommand{\@noticestring}{Technical Report.}
\makeatother
\usepackage[utf8]{inputenc}
\usepackage[T1]{fontenc}
\usepackage{graphicx}
\usepackage{amsmath}
\usepackage{amsfonts}
\usepackage{booktabs}
\usepackage{multirow}
\usepackage{array}
\usepackage{enumitem}
\usepackage{microtype}
\usepackage{xcolor}
\usepackage{url}
\usepackage{xspace}
\usepackage{fancyhdr}
\usepackage{hyperref}

\hypersetup{
    colorlinks=true,
    citecolor=blue!60!black,
    urlcolor=blue!60!black,
    linkcolor=blue!60!black
}

\newcommand{\method}{MiniMind-O\xspace}
\newcommand{\dense}{minimind-3o}
\newcommand{\moe}{minimind-3o-moe}
\newcommand{\githubicon}{\raisebox{-1.5pt}{\includegraphics[height=1.05em]{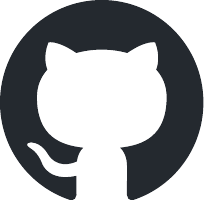}}}
\newcommand{\hficon}{\raisebox{-1.5pt}{\includegraphics[height=1.05em]{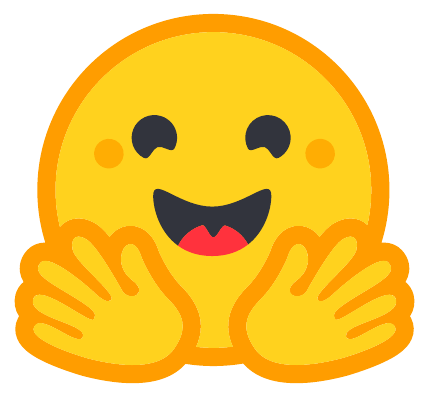}}}
\newcommand{\msicon}{\raisebox{-1.5pt}{\includegraphics[height=0.85em]{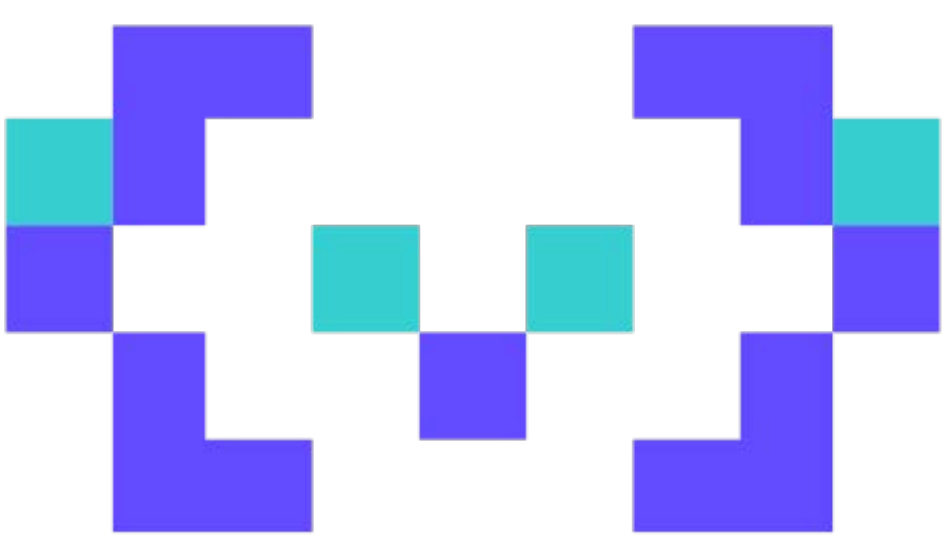}}}

\title{\textbf{MiniMind-O Technical Report: An Open Small-Scale Speech-Native Omni Model}}

\author{%
Jingyao Gong\\
Independent Researcher\\
\texttt{gongjy.cs@foxmail.com}\\[0.6em]
\githubicon\;\href{https://github.com/jingyaogong/minimind-o}{\texttt{github.com/jingyaogong/minimind-o}}\\
\hficon\;\href{https://huggingface.co/collections/jingyaogong/minimind-o}{\texttt{huggingface.co/collections/jingyaogong/minimind-o}}\\
\msicon\;\href{https://www.modelscope.cn/collections/gongjy/minimind-o}{\texttt{modelscope.cn/collections/gongjy/minimind-o}}
}

\fancypagestyle{firstpage}{%
  \fancyhf{}%
  \fancyhead[L]{\includegraphics[height=2.2em]{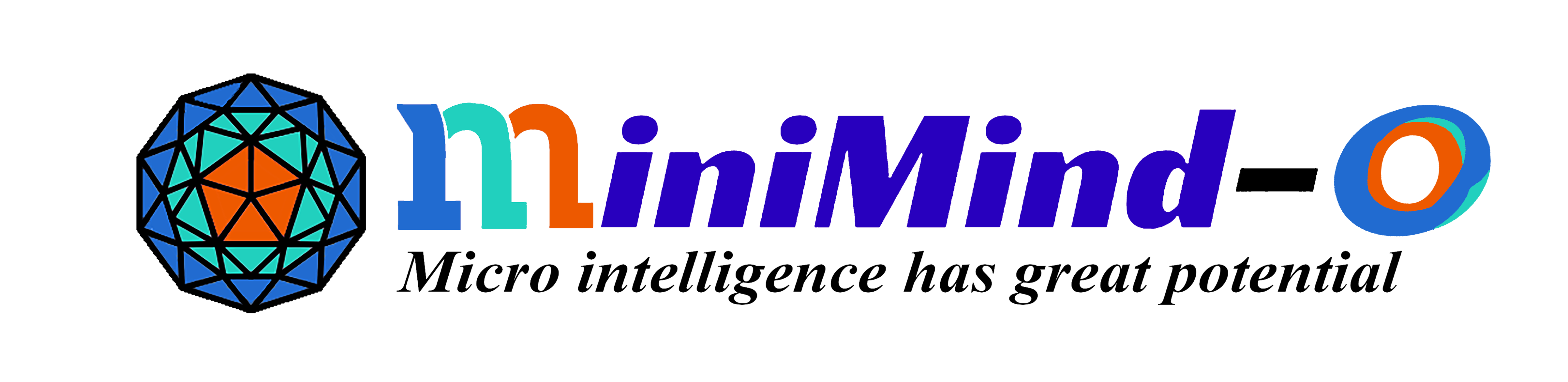}}%
  \fancyhead[R]{}%
}

\begin{document}
\maketitle
\thispagestyle{firstpage}

\begin{abstract}
\method is an open 0.1B-scale omni model built on the MiniMind language model \citep{minimind,minimind-o}. It accepts text, speech, and image inputs, and returns both text and streaming speech. The release includes model code, checkpoints, and the main Parquet training datasets for T2A, I2T, and A2A, making the complete interaction loop directly inspectable. The model uses a full MiniMind backbone as the Thinker and an independent four-layer Talker made from MiniMind blocks. Frozen SenseVoice-Small and SigLIP2 encoders provide speech and image features, which are mapped by lightweight MLP projectors and injected at modality-placeholder positions. The Talker reads a middle-layer Thinker state together with an autoregressive eight-layer Mimi-code buffer. Speaker control is handled by a dedicated speaker token, right-aligned reference codec prompts, and precomputed 192-dimensional CAM++ embeddings, so voice conditioning remains part of the audio-code context rather than a separate TTS module. With a 768-dimensional Talker, the dense and MoE variants reach average CERs of 0.0897 and 0.0900 in Thinker--Talker consistency evaluation, with overall voice-cloning similarities of 0.5995 and 0.5937. Beyond reporting a working system, the paper identifies three scale-critical design choices for small omni models: middle-layer semantic bridging, a released multimodal sequence format, and a parameter-efficient eight-codebook interface.
\end{abstract}

\section{Introduction}

Models such as GPT-4o, Qwen-Omni, Moshi, and recent speech-text systems have moved real-time multimodal interaction from a product interface problem into a model-design problem \citep{gpt4o,moshi,qwen2.5omni,qwen3omni}. A usable system has to listen, see, reason, speak, and stop speaking when the user interrupts. The usual engineering path is still a cascade: ASR turns speech into text, an LLM writes the answer, and TTS renders the waveform. This path works, but it leaves the language model outside the acoustic loop. Once the speech module is external, errors in pronunciation, timing, and speaker control are hard to attribute to a shared representation.

\begin{figure}[!t]
    \centering
    \includegraphics[width=0.99\linewidth]{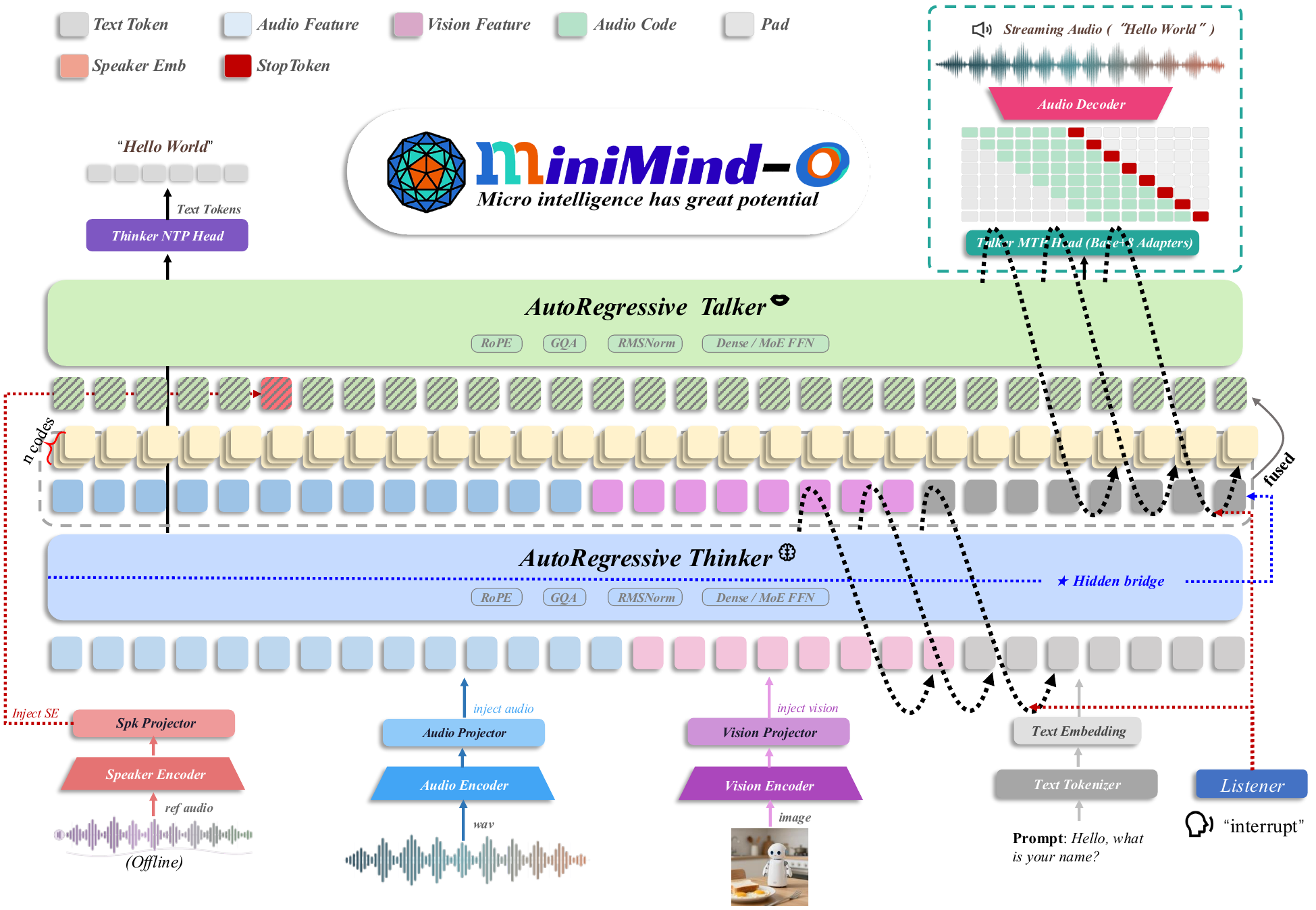}
    \caption{Architecture of \method. Audio and image inputs are encoded by frozen SenseVoice and SigLIP2 encoders, mapped into the MiniMind hidden space by MLP projectors, and injected at modality-placeholder positions. A middle-layer Thinker state is fused with the Mimi-code history by an independent Talker, which predicts eight codec layers for streaming speech generation.}
    \label{fig:architecture}
\end{figure}

\method takes the opposite constraint as the starting point. The base model is MiniMind, not a billion-scale backbone, so every added modality has to pass through a very small hidden space. This makes the system a useful stress test for omni-model design: components that are merely convenient at large scale have to become explicit and measurable at 0.1B scale. The design in Figure~\ref{fig:architecture} keeps the semantic path and the acoustic path separate. The Thinker is the MiniMind transformer itself. It receives normal text embeddings, plus projected SenseVoice and SigLIP2 states injected at audio and image placeholder positions. The Talker is a separate four-layer module initialized from MiniMind blocks when compatible weights are available. This keeps semantic prediction in the language backbone and gives audio-code generation its own recurrent history.

The size is not only a constraint; it is also the main experimental handle. \method{} is intended as a small and fully inspectable omni implementation: it supports text, speech, and image inputs together with streaming speech output while keeping the active model around 0.1B parameters. At this scale, the bridge, projectors, and codec interface have to remain necessary, measurable, and reproducible.

A second result is more architectural. The eight Mimi codebooks could have been given eight independent embedding tables and eight independent output heads. In practice, a non-full-rank shared-base-plus-adapter parameterization gives a clear parameter-efficiency curve: moderate ranks recover most of the convergence and codebook-accuracy gain, while the decoupled rank study shows that the output head rank matters more than the input embedding rank. This makes the low-rank interface an empirically supported design choice rather than an implementation shortcut.

The bridge layer is the third point. If the Talker reads the final next-token-prediction state, it inherits a strong bias toward the current text token and the geometry of the LM head; this is useful for text logits but noisy as an acoustic condition. If it reads too shallow a state, the model has not yet accumulated enough context to resolve pronunciation, syntax, or cross-modal reference. A simple Mandarin example is the character U+5730, whose pronunciation can be \emph{de} or \emph{di} depending on context. A raw embedding does not encode this context-specific pronunciation, while a middle hidden state can carry enough surrounding information without being fully collapsed into the next-token classifier.

The fourth part of the release is the dataset itself. Omni systems are hard to reproduce if the code is open but the alignment data, codec targets, and modality layout are implicit. \method therefore releases the main T2A, I2T, and A2A Parquet datasets together with the code path that consumes them. The dataset is not meant to be a final universal corpus; it is the training substrate for this small-model recipe, with text, image bytes, speech inputs, Mimi code targets, reference-code prompts, and speaker embeddings organized in a format that can be inspected and modified.

The released system has two variants: a dense \dense{} model and a \moe{} model with roughly the same active scale. Audio input is encoded by SenseVoice-Small \citep{sensevoice}; image input is encoded by SigLIP2 \citep{siglip2}; speech output is represented by eight Mimi codebooks and decoded to 24 kHz audio \citep{moshi}. Speaker conditioning is injected by two scale-compatible signals: reference codec prompts and 192-dimensional CAM++ speaker embeddings \citep{campp}. This choice also keeps the inference path inspectable, because no speaker encoder is called inside the model forward pass.

The voice path is therefore closer to in-context conditioning than to a fixed-speaker TTS head. The default release ships five built-in voice prompts, \texttt{dylan}, \texttt{eric}, \texttt{serena}, \texttt{uncle\_fu}, and \texttt{vivian}; an additional seven voices are kept as held-out prompts for evaluation. At inference time, changing the voice only changes the right-aligned reference Mimi codes and the CAM++ vector placed at the \texttt{<|audio\_spk|>} position. The Thinker prompt and Talker weights remain unchanged, which makes voice transfer a property of the shared audio-code layout rather than a separate fine-tuning path.

The report documents the design factors identified as important in this small regime: where to extract the Thinker state, how wide the Talker has to be, how reference speech should be placed in the audio buffer, how the released data is organized, and which evaluation exposes content mismatch rather than only audio quality. These details are not incidental implementation choices. At 0.1B scale, bridge placement, reusable data, and parameter-efficient codebook interfaces directly affect whether the complete loop remains trainable and reproducible. The contribution is therefore not a new large model, but a compact and inspectable recipe that turns speech-native omni interaction into a controllable research object.

\section{Related Work}

\begin{figure}[!t]
    \centering
    \includegraphics[width=0.99\linewidth]{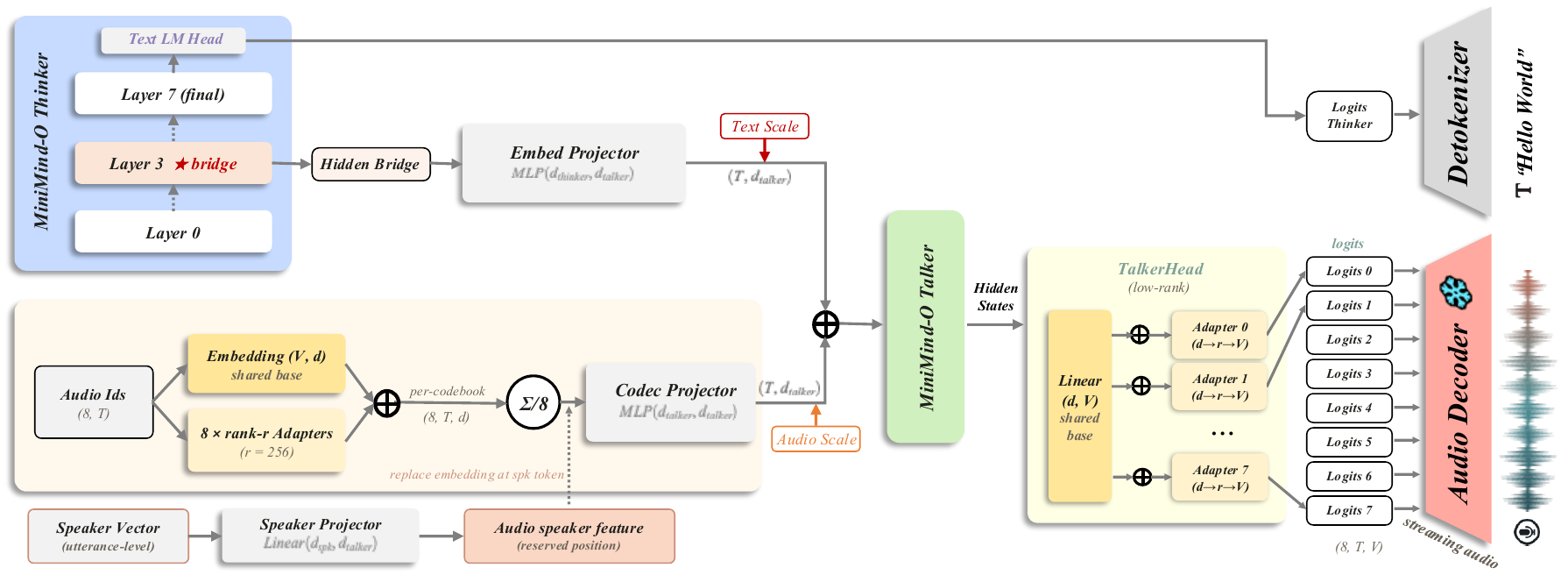}
    \caption{Talker-side speech generation design. The Talker consumes the Thinker bridge state, audio-code embeddings, optional speaker information, and reference codec prompts, then emits eight-layer Mimi codebook logits for waveform decoding.}
    \label{fig:audio-talker}
\end{figure}

\paragraph{Omni and speech-text dialogue models.}
GPT-4o made speech-native multimodal interaction widely visible, and Qwen2.5-Omni and Qwen3-Omni later made the Thinker--Talker recipe more concrete: hidden states can be extracted from a semantic path and consumed by a speech path that runs in streaming mode \citep{gpt4o,qwen2.5omni,qwen3omni}. Open systems explore nearby choices. Mini-Omni showed that speech can be streamed while a language model is still generating text \citep{mini-omni}; Mini-Omni2 added vision and duplex interaction \citep{mini-omni2}. LLaMA-Omni, VITA, GLM-4-Voice, Baichuan-Audio, Step-Audio, and Spirit-LM study related mixtures of speech interaction, audio-language understanding, and interleaved spoken-written modeling \citep{llama-omni,vita,zeng2024glm,li2025baichuan,huang2025step,nguyen2025spirit}. \method uses this line of work as the reference point and studies a complementary question: which components remain necessary when the active model is pushed down to roughly 0.1B parameters, and which interface choices make the resulting loop reproducible rather than only demonstrable.

\paragraph{Discrete audio representation and speech generation.}
Discrete audio tokens are the reason the Talker can be trained with a language-model-style objective. VALL-E showed that codec tokens can carry enough information for zero-shot TTS, MusicGen made multi-codebook autoregression a standard generation pattern, and EnCodec and SNAC provided practical neural codec choices \citep{valle,MusicGEN,encodec,snac}. Moshi introduced Mimi as a streaming audio codec in a speech-text system, while MOSS-Audio-Tokenizer studies scalable tokenizer design for future audio foundation models \citep{moshi,gong2026moss}. \method keeps Mimi's eight-codebook representation. The difference is where the predictor lives: the audio-code predictor is attached to a very small omni model rather than delegated to a large standalone acoustic model.

\paragraph{Multimodal feature alignment.}
For vision-language modeling, CLIP and BLIP-2 established a practical separation between perception and language modeling: a frozen or slowly changing encoder produces features, and a bridge maps them into the LLM space \citep{clip,blip2}. LLaVA, Qwen-VL, Qwen2-VL, and SigLIP2 refine this encoder-side foundation with stronger visual representations and instruction-tuned multimodal use cases \citep{llava,qwenvl,qwen2vl,siglip2}. The MiniMind line has used the same minimal-recipe philosophy in its language-only and vision-language variants \citep{minimind,minimind-v}. In the current \method codebase, both audio and vision use plain two-layer MLP projectors. This is a simpler choice: the external encoders carry perception, and the projectors only have to map their hidden states into the MiniMind embedding space.

\section{Model Architecture}

\begin{figure}[!t]
    \centering
    \includegraphics[width=0.99\linewidth]{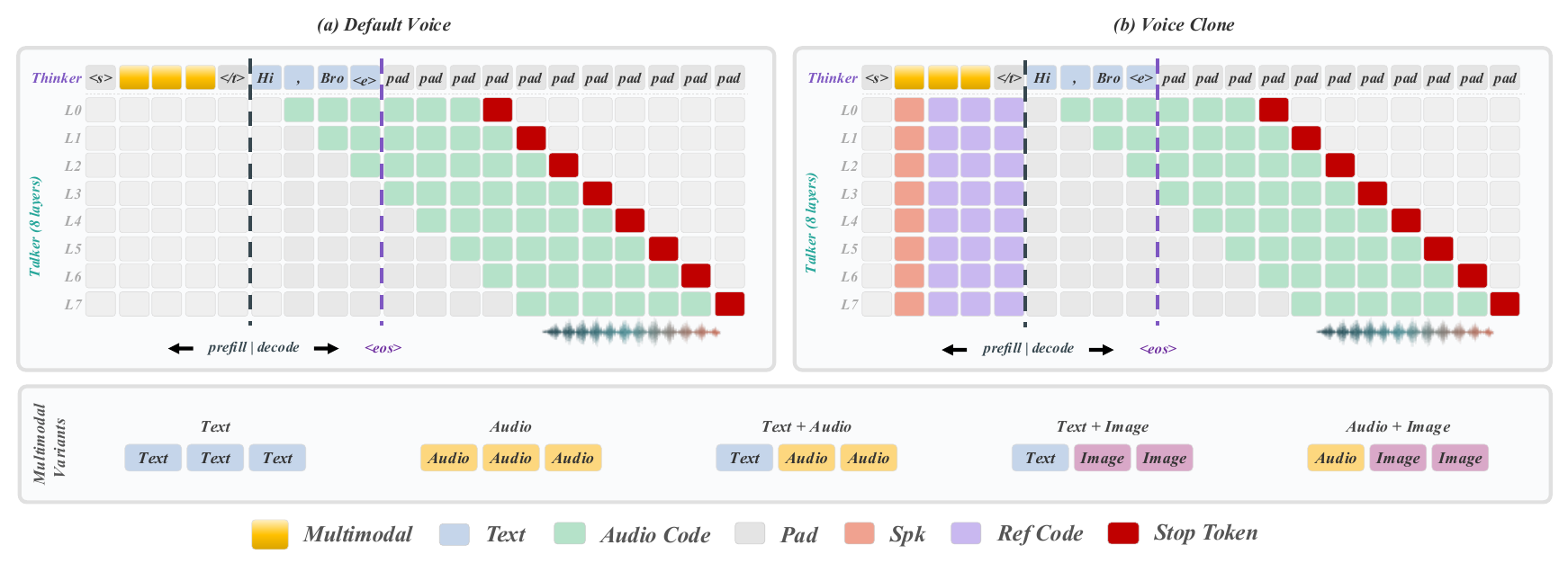}
    \caption{Training sequence format for Thinker and Talker. Text supervision is applied to the Thinker response tokens, while audio supervision is applied to target Mimi code positions. Reference-code regions are used as conditioning context rather than loss targets.}
    \label{fig:sequence-format}
\end{figure}

Figure~\ref{fig:architecture} shows the data path implemented in \texttt{model\_omni.py}. Text enters through the native token embedding table. Speech is converted to SenseVoice frontend features and passed through a frozen SenseVoice encoder; the resulting states are mapped by \texttt{MMAudioProjector}, a two-layer MLP with LayerNorm and GELU. Images are encoded by a frozen SigLIP2 vision model and mapped by the same kind of MLP projector. The projected states preserve the encoder sequence axis and replace contiguous \texttt{<|audio\_pad|>} or \texttt{<|image\_pad|>} embedding positions in the Thinker input sequence.

The Thinker is the full MiniMind transformer, while the Talker is an additional module with \texttt{num\_talker\_hidden\_layers}=4 MiniMind blocks, its own RMSNorm, Mimi-code embedding, codec projection, and audio-code heads. When loading a MiniMind checkpoint that has no Talker weights and the hidden sizes match, the Talker blocks are initialized by copying the last four Thinker blocks. During forward propagation, the Talker input is the sum of two projected streams: \texttt{embed\_proj(bridge\_states)} scaled by a learned text scale, and \texttt{codec\_proj(talker\_emb)} scaled by a learned audio scale. The module therefore reads both semantic states and autoregressive Mimi-code history instead of serving as a simple suffix of the language model.

The audio-code input and output interfaces are intentionally non-full-rank. \texttt{TalkerEmbedding} uses one shared embedding table plus per-codebook low-rank adapters, and \texttt{TalkerHead} uses one shared linear head plus per-codebook low-rank adapters. This compact interface is important at 0.1B scale: the model still sees codebook-specific residuals, while the large shared component is not duplicated eight times.

Figure~\ref{fig:audio-talker} expands the Talker side. Speaker control is represented in the audio-code buffer rather than the text stream. If a speaker embedding is available, the dataset reserves one position before the reference-code region and fills all eight audio layers at that position with \texttt{<|audio\_spk|>}; the model then replaces the Talker embedding at that position with a projected 192-dimensional CAM++ vector. Reference Mimi codes are right-aligned before the target speech region and are masked from the audio loss. This layout makes the reference act as a prompt rather than a reconstruction target, which matters when the same voice has to be reused for a different sentence.

Appendix Table~\ref{tab:modules} lists each module, its concrete model, key configuration, and parameter count. The trainable counts deduplicate the tied MiniMind token embedding and text \texttt{lm\_head}. The evaluation tables keep the experiment-level checkpoint accounting, so they should be read as the model-size labels used for comparison rather than as a decomposition of that table.

\subsection{Middle-layer Bridge}

\begin{figure}[!t]
    \centering
    \includegraphics[width=0.99\linewidth]{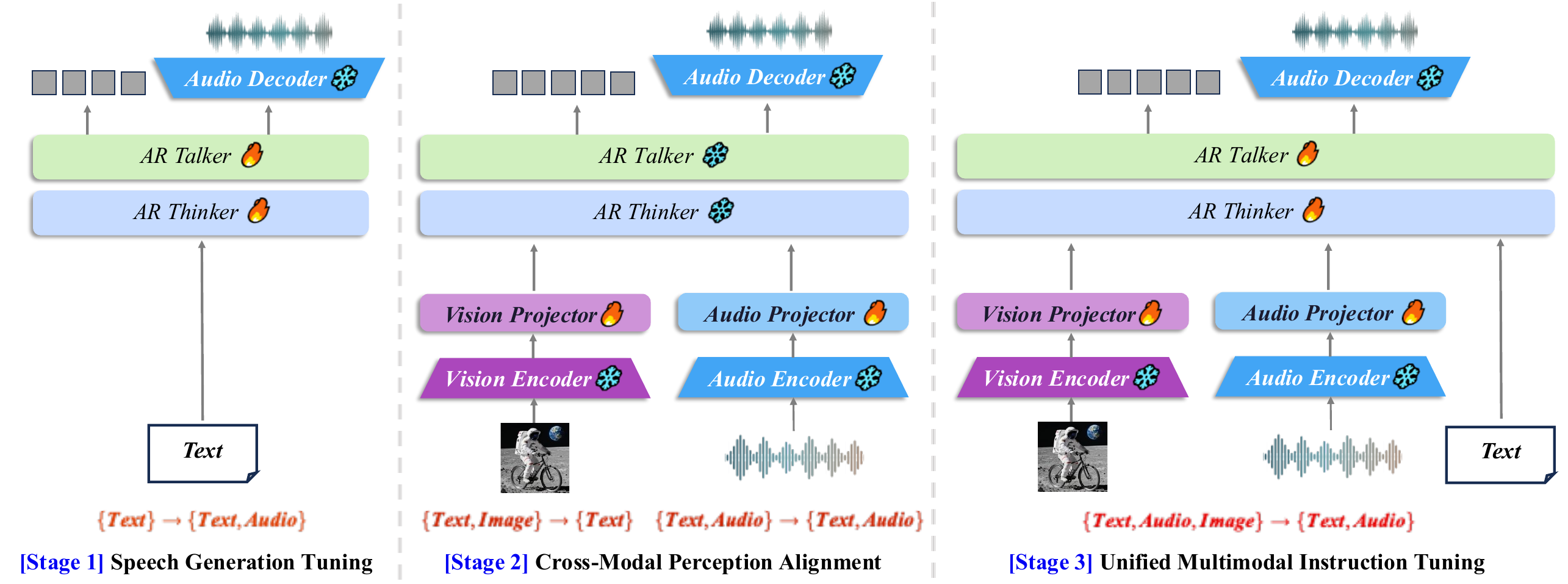}
    \caption{Training pipeline used by the current implementation. The active training script runs \texttt{train\_sft\_omni.py} on T2A, I2T, and A2A data, with \texttt{all} mode for full-model updates and a \texttt{vision\_proj} pass for projector-only visual alignment. SenseVoice and SigLIP2 remain frozen during training.}
    \label{fig:training-pipeline}
\end{figure}

A small omni model is sensitive to the bridge layer. The embedding layer still mainly contains token identity and injected multimodal features; it has not accumulated enough context for pronunciation, syntax, or cross-modal reference. The last layer has the opposite bias. It is already shaped by the next-text-token classifier, so the hidden state carries the geometry and token-selection noise of the LM head rather than the acoustic conditions needed by the Talker. In consistency experiments, moving the bridge too deep increases Talker CER, which is a sign that the acoustic path is being conditioned on states already over-specialized for text logits.

\method therefore extracts the bridge state from a middle Thinker layer, by default \texttt{num\_hidden\_layers // 2 - 1}. The choice is close in spirit to the middle-layer hidden extraction used in Qwen-Omni-style Thinker--Talker systems \citep{qwen2.5omni,qwen3omni}. In the default eight-layer MiniMind setting, this means the bridge is captured after layer 3. A learned \texttt{embed\_proj} maps this state into the Talker hidden space before it is fused with codec-history features. The 768-dimensional Talker is kept because the ablation in Table~\ref{tab:hidden-size} shows that narrower variants lose consistency before the parameter saving becomes useful.

\section{Sequence Format and Streaming Decoding}

\begin{figure}[!t]
    \centering
    \includegraphics[width=0.99\linewidth]{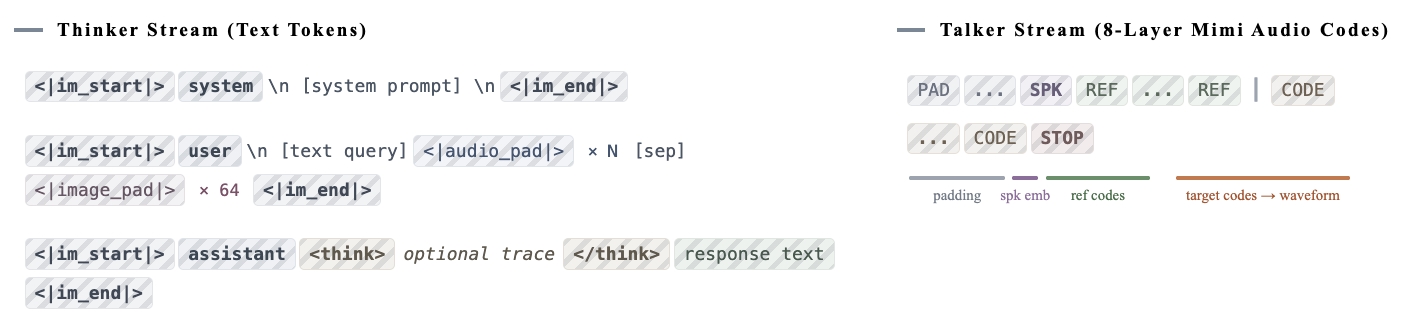}
    \caption{Input token layout in \method. Text tokens, audio placeholders, image placeholders, speaker tokens, reference codes, and target audio codes occupy aligned positions so that the Thinker and Talker can be trained under a single autoregressive schedule.}
    \label{fig:input-token-layout}
\end{figure}

Figure~\ref{fig:sequence-format} and Figure~\ref{fig:input-token-layout} show the actual sequence layout. Each training example is a nine-stream sequence: eight audio-code streams plus one text stream. The Thinker reads the text stream, where repeated audio or image placeholders mark positions to be replaced by projected SenseVoice or SigLIP2 states. The Talker reads the eight audio streams. Before the assistant response, the audio streams are padded, optionally filled with right-aligned reference codes, and optionally marked with a speaker-token position. After the response starts, they carry target Mimi codes. Only the target region receives audio labels; reference and conditioning positions stay masked.

For a response with text tokens \(y_{1:T}\) and Mimi code matrix \(\mathbf{a}\in\mathbb{N}^{8\times T'}\), \method optimizes a joint next-token objective,
\begin{equation}
\mathcal{L} = \mathcal{L}_{\mathrm{text}} + \lambda_{\mathrm{audio}} \sum_{q=1}^{8}\mathcal{L}_{\mathrm{audio}}^{(q)},
\end{equation}
where \(q\) indexes the Mimi codebook layer. Invalid or conditioning-only positions are masked. The dataset staggers audio targets by codebook layer: layer \(q\) starts at \texttt{assistant\_start + q + 1}. In streaming inference, the first generated text step has no audio output, and the eight codec layers become available with the same delayed schedule. Once a complete eight-layer frame is available, the Mimi codes can be decoded incrementally into 24 kHz waveform, so playback can begin before the full textual response is complete.

This format makes the evaluation stricter than a cascaded ASR--LLM--TTS system in one specific sense. The Talker is judged against the Thinker's own text, not against an external transcript or a hand-written reference. When numerals, rare names, or longer clauses are not rendered correctly, the mismatch can be traced back to the shared omni path. A large standalone TTS module may absorb part of this difficulty; here the behavior remains visible.

\section{Training Pipeline}

The current training entry is \texttt{train\_sft\_omni.py}. Its mode switch is deliberately small: \texttt{all} updates the trainable MiniMind/Talker/projector parameters together, \texttt{audio\_proj} freezes the rest of the model and trains only the audio projector, and \texttt{vision\_proj} does the same for the vision projector. The active \texttt{train.sh} runs full-model passes over \texttt{sft\_t2a}, \texttt{sft\_i2t}, and \texttt{sft\_a2a}, followed by a projector-only \texttt{sft\_i2t} pass, for both dense and MoE variants. This differs from the older README description that names separate \texttt{t2t}, \texttt{t2a}, and \texttt{a2a} modes; those names describe the data type, not the current command-line mode interface.

All runs reported in this paper are produced on a single workstation with four NVIDIA RTX 3090 GPUs (24\,GB each), using PyTorch DDP launched via \texttt{torchrun --nproc\_per\_node 4}. Training uses bf16 mixed precision with the AdamW optimizer, a per-GPU batch size of 32, no gradient accumulation, and gradient clipping at 1.0. The full-model T2A pass uses learning rate $5\times10^{-6}$ for one epoch on \texttt{sft\_t2a}; the audio-projector A2A pass and the vision-projector I2T pass use $5\times10^{-4}$ and $5\times10^{-5}$ respectively for one epoch each; the full-model A2A pass uses $5\times10^{-5}$ for three epochs on \texttt{sft\_a2a}; and the full-model I2T pass uses $5\times10^{-6}$ for one epoch with a 768-token context. Wall-clock time per stage is approximately 45\,min for T2A, 25\,min for the audio-projector A2A pass, 75\,min for the three-epoch A2A pass, and 45\,min for each I2T pass, so a complete dense or MoE training cycle finishes in under four hours on this setup. Working at 0.1B active scale is what makes this consumer-GPU schedule feasible: at frontier scale, the same loop would not be reproducible without a much larger compute budget.

\begin{table}[t]
\centering
\caption{Main training datasets used by \method. Audio durations are computed from the pre-extracted Mimi-code statistics in the released dataset.}
\label{tab:data}
\small
\begin{tabular}{lrrrr}
\toprule
Dataset & Items & Input speech & Output speech & Total speech \\
\midrule
\texttt{sft\_i2t} & \(\sim\)100K & -- & -- & -- \\
\texttt{sft\_t2a} & 1,248,923 & -- & 1636.01 h & 1636.01 h \\
\texttt{sft\_a2a} & 414,024 & 1711.97 h & 423.40 h & 2135.37 h \\
\bottomrule
\end{tabular}
\end{table}

Table~\ref{tab:data} gives the data scale used in the release. The public dataset is part of the contribution because it fixes the exact sequence and codec layout used by the model rather than leaving reproduction to a private preprocessing pipeline. \texttt{sft\_t2a} contains 1,248,923 samples and 1636.01 h of output speech. \texttt{sft\_a2a} contains 414,024 samples, 1711.97 h of input speech, and 423.40 h of output speech. The text-to-audio split is close to balanced between Chinese and English outputs, with 45.7\% Chinese, 46.5\% English, and 7.8\% mixed content. The audio-to-audio split is Chinese-heavy: 70.8\% Chinese, 21.2\% English, and 8.0\% mixed content. This distribution shows up in behavior. Short Chinese and English replies are usually stable; longer English speech is where pronunciation drift and omissions become easier to trigger.

\begin{figure}[!t]
    \centering
    \includegraphics[width=0.99\linewidth]{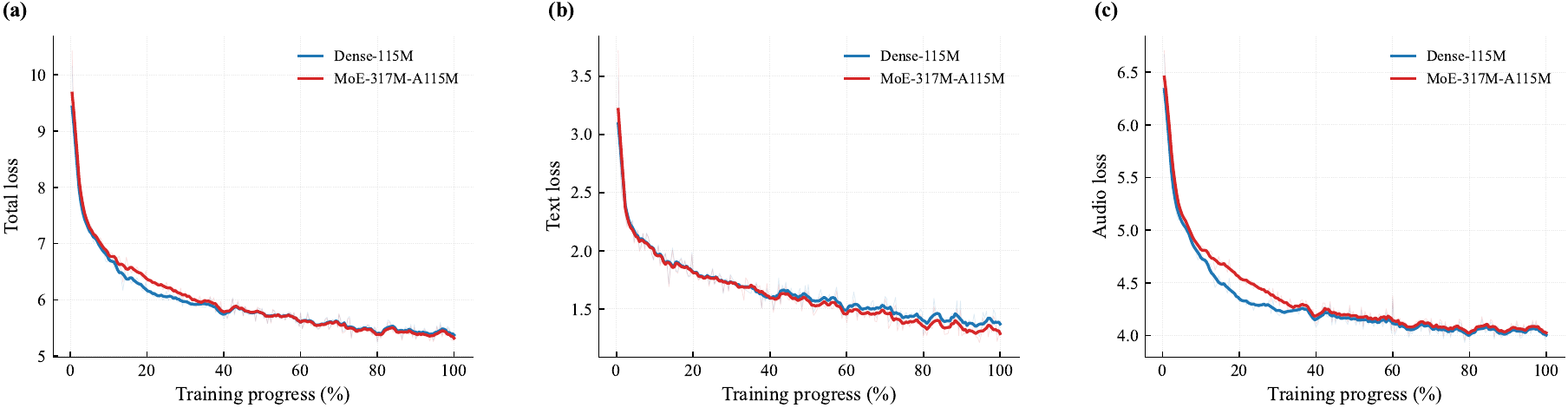}
    \caption{Text-to-audio training curves for \dense{} and \moe{}. The plotted curve uses the cleaned log segment after removing the erroneous resume interval caused by loading an incompatible checkpoint.}
    \label{fig:t2a-curves}
\end{figure}

\begin{figure}[!t]
    \centering
    \includegraphics[width=0.99\linewidth]{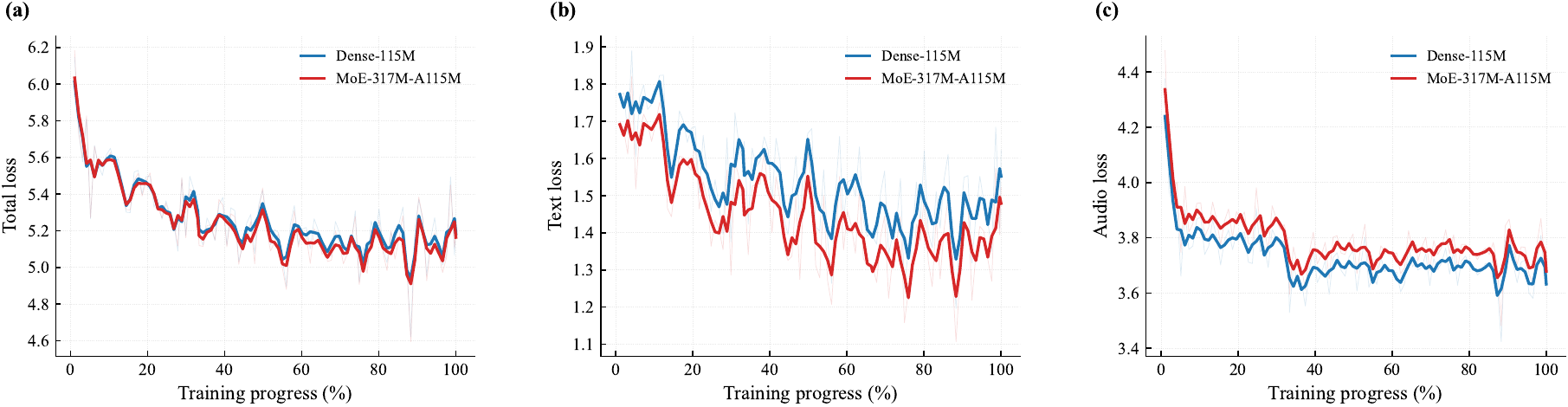}
    \caption{Audio-to-audio training curves for \dense{} and \moe{}. The A2A stage is trained after text-to-audio learning and exposes the full speech-in/speech-out loop.}
    \label{fig:a2a-curves}
\end{figure}

Figure~\ref{fig:t2a-curves} and Figure~\ref{fig:a2a-curves} show the two speech-generation stages. The T2A curve uses the cleaned log segment; an earlier resume from an incompatible checkpoint produced a loss spike, and that interval is not used here. The MoE variant has a larger total parameter count but roughly the same active scale as the dense model, so these curves are more useful for reading capacity allocation than for claiming equal-compute superiority.

\begin{figure}[!t]
    \centering
    \includegraphics[width=0.99\linewidth]{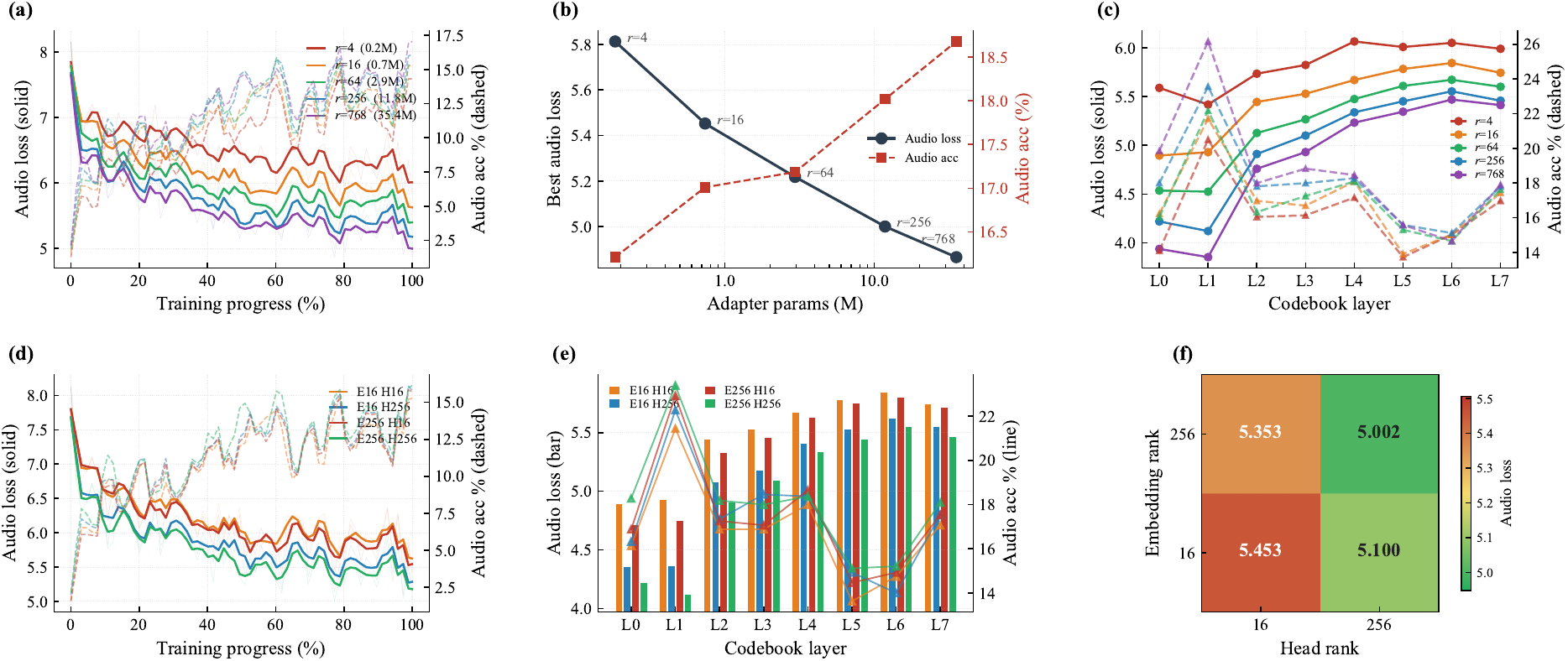}
    \caption{Rank ablation for the Talker-side low-rank interfaces. The top row sweeps a unified rank for \texttt{TalkerEmbedding} and \texttt{TalkerHead}; the bottom row decouples the two ranks. Solid curves or bars report audio loss, while dashed curves or overlaid markers report audio accuracy. The results show that moderate ranks already recover most of the parameter-efficient gain, and that the output head rank is more important than the embedding rank.}
    \label{fig:rank-ablation}
\end{figure}

Figure~\ref{fig:rank-ablation} isolates the Talker-side low-rank interfaces from the rest of the model. The experiment freezes the Thinker and varies the rank of the \texttt{TalkerEmbedding} and \texttt{TalkerHead} adapters on the same A2A subset. Increasing the unified rank improves convergence, final audio loss, and codebook accuracy, but the gain becomes gradual once the adapter reaches a few million parameters. The decoupled runs are more diagnostic: increasing the \texttt{TalkerHead} rank from 16 to 256 gives a larger improvement than increasing the \texttt{TalkerEmbedding} rank under the same setting. This matches the roles of the two interfaces. The embedding side mainly reads recent Mimi-code history, while the head side has to separate eight codebook distributions over the full audio vocabulary.

\section{Evaluation}
\label{sec:evaluation}

The evaluation is built around consistency properties that are easy to miss in demos. For each prompt, the model produces Thinker text and Talker audio. The audio is transcribed by Qwen3-ASR-Flash, and the transcript is compared with the Thinker text. The internal consistency runs report CER, while the cross-model English and vision-language comparisons additionally report WER. These metrics leave naturalness and preference to separate evaluation; here they ask a narrower question: after the Talker turns the hidden state into waveform, does the spoken or written output still match the intended text? The protocol is therefore ASR-dependent and should not be read as a MOS or preference study. In particular, numeral formatting can inflate edit distance when the waveform is correct but the ASR writes a number in words.

\begin{table*}[t]
\centering
\caption{Talker hidden-size ablation. The 768-dimensional Talker is selected for both variants because it gives the best average CER and keeps the Thinker--Talker dimensional interface simple.}
\label{tab:hidden-size}
\small
\begin{tabular}{llrrrr}
\toprule
Variant & Talker hidden & Params & Avg CER $\downarrow$ & Short $\downarrow$ & Mid / Long $\downarrow$ \\
\midrule
Dense & 768 & 115.29M & 0.0897 & 0.1528 & 0.0874 / 0.0675 \\
Dense & 512 & 96.13M & 0.1745 & 0.2709 & 0.2455 / 0.0976 \\
Dense & 384 & 88.72M & 0.2767 & 0.3904 & 0.1865 / 0.4046 \\
\midrule
MoE & 768 & 317.05M-A115.33M & 0.0900 & 0.2075 & 0.0533 / 0.0271 \\
MoE & 512 & 261.32M-A96.17M & 0.1265 & 0.0711 & 0.1490 / 0.1464 \\
MoE & 384 & 240.04M-A88.75M & 0.3280 & 0.3757 & 0.2777 / 0.4313 \\
\bottomrule
\end{tabular}
\end{table*}

Table~\ref{tab:hidden-size} reports the Talker hidden-size ablation. The 768-dimensional setting is the only one that stays stable for both dense and MoE variants. Reducing the Talker to 512 or 384 does save parameters, but it also narrows the acoustic state seen by each codebook head. Since Mimi prediction is an eight-layer problem, the bottleneck is amplified across codebooks. The ablation rules out a simple scaling assumption: the Talker cannot be made very thin just because the semantic plan comes from the Thinker.

\begin{table}[t]
\centering
\caption{Voice-cloning similarity measured by CAM++ speaker embeddings \citep{campp}. The baseline row refers to the earlier reference-code-only setting reported during development.}
\label{tab:voice-clone}
\small
\begin{tabular}{lrrr}
\toprule
Model & Seen $\uparrow$ & Unseen $\uparrow$ & Overall $\uparrow$ \\
\midrule
Previous baseline & 0.6150 & 0.5310 & -- \\
\dense{} & 0.6472 & 0.5654 & 0.5995 \\
\moe{} & 0.6267 & 0.5702 & 0.5937 \\
\bottomrule
\end{tabular}
\end{table}

Table~\ref{tab:voice-clone} shows the voice-cloning evaluation; the per-speaker breakdown is in Appendix Table~\ref{tab:voice-clone-speaker}. The seen split uses the five built-in voices shipped in \texttt{voices.pt}: \texttt{dylan}, \texttt{eric}, \texttt{serena}, \texttt{uncle\_fu}, and \texttt{vivian}. The unseen split uses seven prompts from \texttt{voices\_unseen.pt}: \texttt{arthur}, \texttt{chelsie}, \texttt{cherry}, \texttt{ethan}, \texttt{jennifer}, \texttt{momo}, and \texttt{moon}. For each voice, generation keeps the same textual questions and changes only the in-context speaker condition, namely the reference Mimi codes and the 192-dimensional CAM++ vector. Dense is slightly better on seen speakers, and MoE is slightly better on unseen speakers, but the overall gap is small. Both improve over the earlier reference-code-only baseline, from 0.6150 to 0.6472 on seen voices for the dense model and from 0.5310 to 0.5702 on unseen voices for the MoE model. The per-speaker table shows that the best individual voices (\texttt{uncle\_fu}, \texttt{serena}, \texttt{arthur}) exceed 0.70 cosine similarity for at least one variant, while the low outliers (\texttt{eric} under MoE, \texttt{moon} under dense) usually coincide with degraded generated audio before the speaker encoder is applied.

\begin{table}[t]
\centering
\caption{Cross-model English T2A consistency under the same brief-answer constraint. \dense{} is smaller than Mini-Omni and Mini-Omni2 \citep{mini-omni,mini-omni2}, but the gap is concentrated in medium-length answers.}
\label{tab:cross-t2a}
\small
\begin{tabular}{lrrr}
\toprule
Model & Params & Avg CER $\downarrow$ & Avg WER $\downarrow$ \\
\midrule
Mini-Omni & 0.5B & 0.0101 & 0.0185 \\
Mini-Omni2 & 0.5B & 0.0371 & 0.0431 \\
\dense{} & 0.1B & 0.0964 & 0.0973 \\
\bottomrule
\end{tabular}
\end{table}

\begin{table}[t]
\centering
\caption{Vision-language comparison with length-matched references generated by Qwen-VL-Plus \citep{qwenvl}. CER/WER are high because open-ended image descriptions admit many valid paraphrases.}
\label{tab:vl}
\small
\begin{tabular}{lrrr}
\toprule
Model & Params & Avg CER $\downarrow$ & Avg WER $\downarrow$ \\
\midrule
Mini-Omni2 & 0.5B & 0.7609 & 0.9756 \\
\dense{} & 0.1B & 0.8241 & 1.0293 \\
\bottomrule
\end{tabular}
\end{table}

Table~\ref{tab:vl} reports a small vision-language comparison. Mini-Omni does not support this path, so the comparison is between Mini-Omni2 and \dense{} \citep{mini-omni,mini-omni2}. The evaluation uses nine synthetic images; for each output, Qwen-VL-Plus generates a separate length-matched reference \citep{qwenvl}. The absolute values are high because open-ended image descriptions admit many valid paraphrases. Under the same protocol, \dense{} trails Mini-Omni2 but remains in the same order of magnitude while using about one fifth of the parameters. Per-sample values are in Appendix Table~\ref{tab:vl-detail}.

\section{Discussion and Limitations}

The main lesson from \method{} is that the omni loop has a meaningful small-model regime. A full text--speech--image loop can be made public and inspectable at roughly 0.1B active parameters; the training data can be released in a form that preserves the actual multimodal layout; the eight-codebook embedding/head interface does not have to be fully duplicated across codebooks; and a middle-layer bridge gives the Talker a cleaner semantic condition than the final next-token-prediction state. These are positive results even though the model remains far from frontier-scale systems.

The limitations are also clear. Speech naturalness and long-form stability remain behind larger speech-text models, with medium-length English answers being the most visible weak point. The visual pathway uses a frozen SigLIP2 encoder, 64 placeholder positions, and a plain MLP projector, so its role is closer to a compact vision-to-speech path than to a large-VLM replacement. Voice cloning improves over the earlier reference-code-only baseline, while still depending heavily on reference quality and on whether the generated audio is clean enough for the speaker encoder to read. The MoE variant is best read as a capacity-allocation experiment rather than a final expert layout. The evaluation is also deliberately narrow: the main automatic scores measure transcript consistency, not human naturalness, latency under load, safety behavior, or robustness to noisy far-field speech.

The claim is intentionally narrow. \method{} is not presented as a competitor to frontier-scale systems; its value is that the complete omni loop can be reproduced and inspected without hiding the key choices behind scale.

\section{Conclusion}

This report introduced \method, a 0.1B-scale open omni model with text, speech, and image inputs and streaming speech output. The current code combines a full MiniMind Thinker, an independent four-layer Talker, middle-layer semantic bridging, MLP-based audio/vision projection, Mimi-code speech generation, and staged SFT over released T2A, I2T, and A2A data. The dense and MoE variants both maintain usable Thinker--Talker consistency under short-answer settings, support speaker-conditioned generation, and run basic vision-language-to-speech interaction. The broader message is that small omni models can serve as controlled research artifacts: with public data, a middle hidden bridge, and low-rank codebook-specific embedding/head adapters, a complete loop can be made parameter-efficient enough to study directly. In this sense, \method contributes a reproducible small-scale baseline for analyzing speech-native omni design, not only a runnable demo. The remaining gaps are exposed by the same implementation, which makes the small regime useful for analysis rather than only for deployment efficiency.

\bibliographystyle{plainnat}
\bibliography{ref}

\clearpage
\appendix
\makeatletter
\renewcommand\section{\@startsection{section}{1}{\z@}%
  {-2.5ex \@plus -1ex \@minus -.2ex}%
  {1.5ex \@plus .2ex}%
  {\normalfont\large\bfseries}}
\renewcommand{\@seccntformat}[1]{\csname the#1\endcsname.\quad}
\makeatother
\renewcommand{\thesection}{\Alph{section}}

{\Large\bfseries Appendices}\par\vspace{1.5em}

\section{Module and Evaluation Details}
\label{app:eval-detail}

This appendix collects the detailed tables referenced in the main text. Table~\ref{tab:modules} enumerates every module in the current \method{} implementation together with its concrete model, key hyperparameters, and parameter count. The trainable counts deduplicate the tied MiniMind token embedding and text \texttt{lm\_head}; frozen modules are loaded as-is and never updated during training.

\begin{table*}[!ht]
\centering
\caption{Main modules used by the current implementation. Trainable component counts are taken from the current PyTorch modules; external perception and codec models are frozen and are not counted as active \method{} parameters.}
\label{tab:modules}
\begin{tabular}{@{}p{0.15\linewidth}p{0.22\linewidth}p{0.35\linewidth}p{0.20\linewidth}@{}}
\toprule
Module & Concrete model or layer & Key configuration & Status / params (Dense / MoE) \\
\midrule
Thinker & MiniMind Transformer & 8 layers, hidden 768, 8 query heads, 4 KV heads, vocab 6400 & trainable, 63.91M / 198.42M \\
Talker & independent MiniMind blocks & 4 layers, hidden 768, audio vocab 2112, 8 codebook heads, rank-256 embedding/head adapters & trainable, 47.05M / 114.30M \\
Audio projector & \texttt{MMAudioProjector} & LayerNorm(512) -- Linear -- GELU -- Linear to hidden 768 & trainable, 0.99M \\
Vision projector & \texttt{MMVisionProjector} & LayerNorm(768) -- Linear -- GELU -- Linear to hidden 768 & trainable, 1.18M \\
Audio encoder & SenseVoice-Small & 50 encoder blocks, output size 512, 16 kHz frontend & frozen, 234.00M \\
Vision encoder & SigLIP2 base patch32-256 & 12 layers, hidden 768, 12 heads, 64 image tokens & frozen, 94.55M \\
Speech codec & Mimi & 8 codebooks, size 2048, 12.5 Hz frames, 24 kHz waveform & frozen, 96.15M \\
Speaker condition & CAM++ embedding & 192-dimensional vector projected by \texttt{spk\_proj} & precomputed, no online encoder \\
\bottomrule
\end{tabular}
\end{table*}

Table~\ref{tab:voice-clone-speaker} breaks down voice-cloning similarity by individual speaker. The five seen voices are the built-in prompts shipped in \texttt{voices.pt}; the seven unseen voices come from \texttt{voices\_unseen.pt} and are never seen during training. For each voice the same set of textual questions is used, changing only the in-context speaker condition (reference Mimi codes and 192-dimensional CAM++ vector). The best individual voices (\texttt{uncle\_fu}, \texttt{serena}, \texttt{arthur}) exceed 0.70 cosine similarity for at least one variant, while the lowest outliers (\texttt{eric} under \moe{}, \texttt{moon} under \dense{}) typically coincide with degraded generated audio quality before the speaker encoder is applied.

\begin{table*}[!ht]
\centering
\caption{Per-speaker voice-cloning similarity measured by CAM++ cosine similarity. The seen speakers are the built-in prompts shipped with the release; unseen speakers are held out from the default voice set.}
\label{tab:voice-clone-speaker}
\begin{tabular}{llrr}
\toprule
Split & Speaker & \dense{} $\uparrow$ & \moe{} $\uparrow$ \\
\midrule
Seen & \texttt{dylan} & 0.6997 & 0.6837 \\
Seen & \texttt{eric} & 0.5289 & 0.4232 \\
Seen & \texttt{serena} & 0.7092 & 0.7041 \\
Seen & \texttt{uncle\_fu} & 0.7241 & 0.7337 \\
Seen & \texttt{vivian} & 0.5744 & 0.5888 \\
\midrule
Unseen & \texttt{arthur} & 0.7171 & 0.6750 \\
Unseen & \texttt{chelsie} & 0.6437 & 0.6240 \\
Unseen & \texttt{cherry} & 0.5689 & 0.5678 \\
Unseen & \texttt{ethan} & 0.4783 & 0.4847 \\
Unseen & \texttt{jennifer} & 0.4749 & 0.4003 \\
Unseen & \texttt{momo} & 0.6470 & 0.5720 \\
Unseen & \texttt{moon} & 0.4282 & 0.6673 \\
\bottomrule
\end{tabular}
\end{table*}

Tables~\ref{tab:cross-t2a-bucket} and~\ref{tab:cross-t2a-detail} expand the cross-model English T2A comparison from the main text. All three models receive the same instruction (\texttt{Answer briefly in one short sentence}). The length-bucket view (Table~\ref{tab:cross-t2a-bucket}) shows that \dense{} is competitive with Mini-Omni2 on short answers ($\leq$15 words) but falls behind on medium-length responses (16--30 words), where the Talker must sustain pronunciation and lexical consistency across a full clause.

\begin{table*}[!ht]
\centering
\caption{Length-bucket breakdown for the cross-model English T2A comparison. Each entry reports CER / WER with the number of evaluated samples in parentheses.}
\label{tab:cross-t2a-bucket}
\begin{tabular}{lccc}
\toprule
Length bucket & Mini-Omni & Mini-Omni2 & \dense{} \\
\midrule
Short (\(\leq 15\) words) & 0.0195 / 0.0384 (n=8) & 0.0503 / 0.0584 (n=14) & 0.0531 / 0.0417 (n=8) \\
Mid (16--30 words) & 0.0038 / 0.0052 (n=12) & 0.0062 / 0.0076 (n=6) & 0.1327 / 0.1420 (n=11) \\
Long (31--60 words) & -- & -- & 0.0431 / 0.0508 (n=1) \\
\bottomrule
\end{tabular}
\end{table*}

The per-question breakdown (Table~\ref{tab:cross-t2a-detail}) reveals that 14 out of 20 questions achieve zero CER for all three models. The few high-CER outliers are mainly driven by surface-form mismatches rather than clear pronunciation failures. For example, question 04 involves the number ``299,792,458'', while the ASR may transcribe the spoken answer as ``two hundred ninety-nine million\ldots'', inflating character-level distance. Question 13 shows the same metric sensitivity for named entities, where a small transcript variation can dominate the score for a short answer.

\begin{table*}[!ht]
\centering
\caption{Per-question cross-model English T2A comparison. Each cell reports CER / WER. Questions are abbreviated; all are prefixed with ``Answer briefly in one short sentence.'' Entries with CER $>$ 0.3 are typically caused by surface-form ASR mismatches such as number spelling or named-entity variants.}
\label{tab:cross-t2a-detail}
\small
\begin{tabular}{@{}rlccc@{}}
\toprule
\# & Question (abbreviated) & Mini-Omni & Mini-Omni2 & \dense{} \\
\midrule
00 & Hello, how are you today? & 0.000 / 0.000 & 0.000 / 0.000 & 0.000 / 0.000 \\
01 & Can you tell me a joke? & 0.000 / 0.000 & 0.000 / 0.000 & 0.000 / 0.000 \\
02 & What is the capital of France? & 0.000 / 0.000 & 0.000 / 0.000 & 0.000 / 0.000 \\
03 & How do you make a cup of coffee? & 0.000 / 0.000 & 0.000 / 0.000 & 0.000 / 0.000 \\
04 & What is the speed of light? & 0.000 / 0.000 & 0.382 / 0.286 & 1.410 / 1.471 \\
05 & Explain what AI is. & 0.000 / 0.000 & 0.000 / 0.000 & 0.000 / 0.000 \\
06 & Tallest mountain in the world? & 0.000 / 0.000 & 0.000 / 0.000 & 0.000 / 0.000 \\
07 & How many planets in the solar system? & 0.156 / 0.125 & 0.303 / 0.250 & 0.000 / 0.000 \\
08 & What causes rainbows? & 0.000 / 0.000 & 0.000 / 0.000 & 0.000 / 0.000 \\
09 & Recommend a good book? & 0.000 / 0.182 & 0.000 / 0.182 & 0.000 / 0.000 \\
10 & Largest ocean on Earth? & 0.000 / 0.000 & 0.000 / 0.000 & 0.000 / 0.000 \\
11 & How does photosynthesis work? & 0.000 / 0.000 & 0.000 / 0.000 & 0.017 / 0.045 \\
12 & Benefits of regular exercise? & 0.000 / 0.000 & 0.000 / 0.000 & 0.000 / 0.000 \\
13 & Who invented the telephone? & 0.000 / 0.000 & 0.000 / 0.000 & 0.425 / 0.333 \\
14 & Meaning of life? & 0.000 / 0.000 & 0.000 / 0.000 & 0.000 / 0.000 \\
15 & How do airplanes stay in the air? & 0.000 / 0.000 & 0.019 / 0.100 & 0.033 / 0.045 \\
16 & Virus vs.\ bacteria? & 0.000 / 0.000 & 0.000 / 0.000 & 0.000 / 0.000 \\
17 & Explain blockchain technology. & 0.000 / 0.000 & 0.000 / 0.000 & 0.000 / 0.000 \\
18 & Three tips for better sleep? & 0.045 / 0.063 & 0.037 / 0.045 & 0.043 / 0.051 \\
19 & Why is the sky blue? & 0.000 / 0.000 & 0.000 / 0.000 & 0.000 / 0.000 \\
\midrule
\multicolumn{2}{@{}l}{Average} & 0.010 / 0.019 & 0.037 / 0.043 & 0.096 / 0.097 \\
\bottomrule
\end{tabular}
\end{table*}

Table~\ref{tab:vl-detail} gives the per-sample vision-language results. Mini-Omni does not support this path, so the comparison is limited to Mini-Omni2 and \dense{}. Each image is described independently; Qwen-VL-Plus generates a separate length-matched reference for the same image, and CER/WER are computed against that reference. The absolute values are high across both models because open-ended image descriptions admit many valid paraphrases and detail orderings---two correct descriptions of the same image can share very few exact n-grams. Under this protocol \dense{} trails Mini-Omni2 but stays within the same range while using about one fifth of the parameters.

\begin{table*}[!ht]
\centering
\caption{Per-sample vision-language comparison. Each cell reports output length / reference length / CER / WER.}
\label{tab:vl-detail}
\begin{tabular}{rcc}
\toprule
\# & Mini-Omni2 & \dense{} \\
\midrule
00 & 158 / 128 / 0.7976 / 1.0391 & 111 / 103 / 0.7883 / 0.9709 \\
01 & 260 / 233 / 0.7273 / 0.9914 & 109 / 87 / 0.8013 / 1.0920 \\
02 & 110 / 89 / 0.7957 / 0.9888 & 126 / 98 / 0.8728 / 1.0816 \\
03 & 79 / 84 / 0.7408 / 0.9286 & 113 / 104 / 0.8031 / 0.9519 \\
04 & 94 / 102 / 0.7273 / 0.8529 & 130 / 101 / 0.8531 / 1.0594 \\
05 & 91 / 72 / 0.7209 / 1.0556 & 115 / 91 / 0.8629 / 1.0989 \\
06 & 187 / 194 / 0.7293 / 0.9021 & 119 / 118 / 0.7551 / 0.8814 \\
07 & 295 / 267 / 0.7682 / 0.9625 & 114 / 89 / 0.8548 / 1.0449 \\
08 & 143 / 118 / 0.8414 / 1.0593 & 137 / 109 / 0.8254 / 1.0826 \\
\bottomrule
\end{tabular}
\end{table*}

\newpage

\section{Qualitative Examples}
\label{app:qualitative}

\begin{figure}[!ht]
    \centering
    \includegraphics[width=0.99\linewidth]{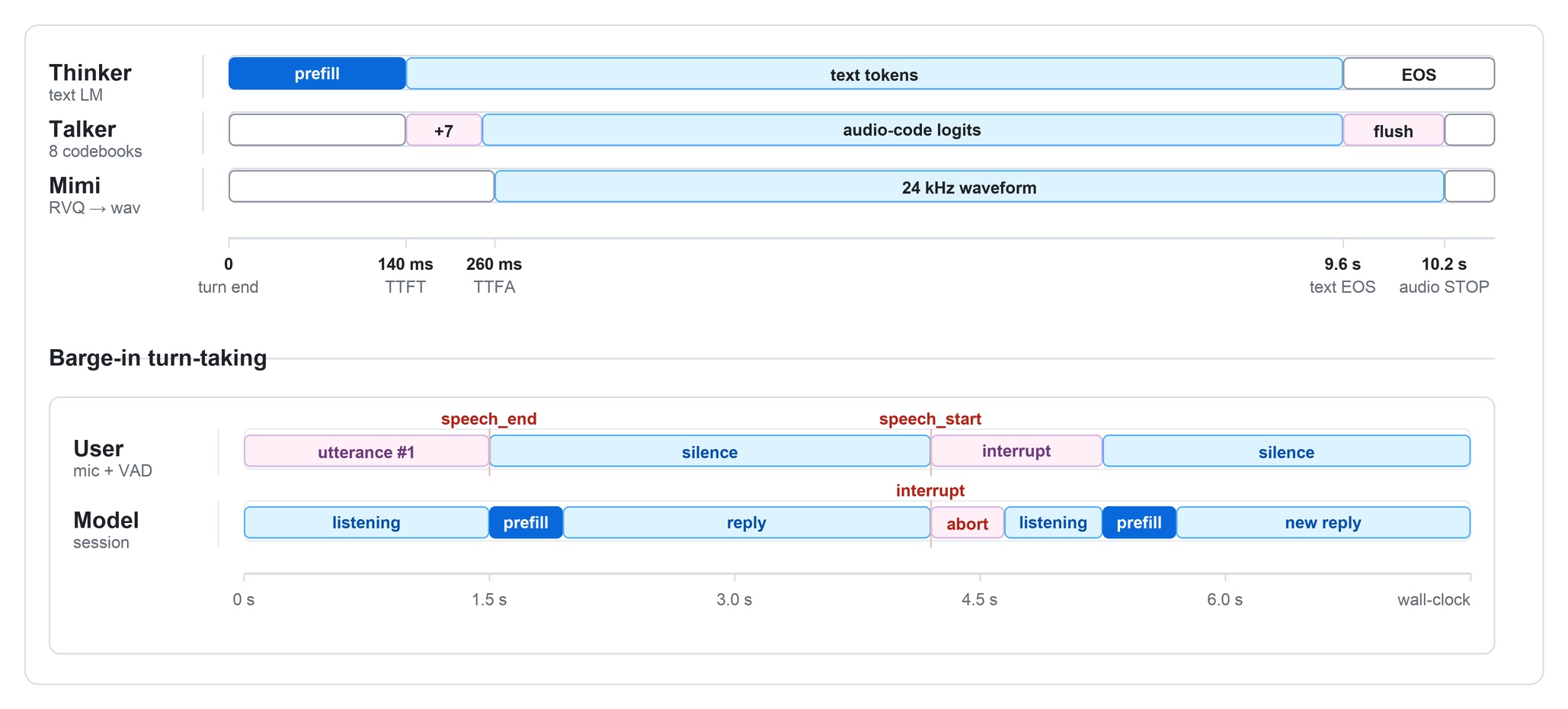}
    \caption{Real-time interaction interface. Streaming speech generation allows playback while decoding continues, and VAD-triggered barge-in can stop the current output when a new user turn is detected.}
    \label{fig:realtime}
\end{figure}

\begin{figure}[!ht]
    \centering
    \includegraphics[width=0.99\linewidth]{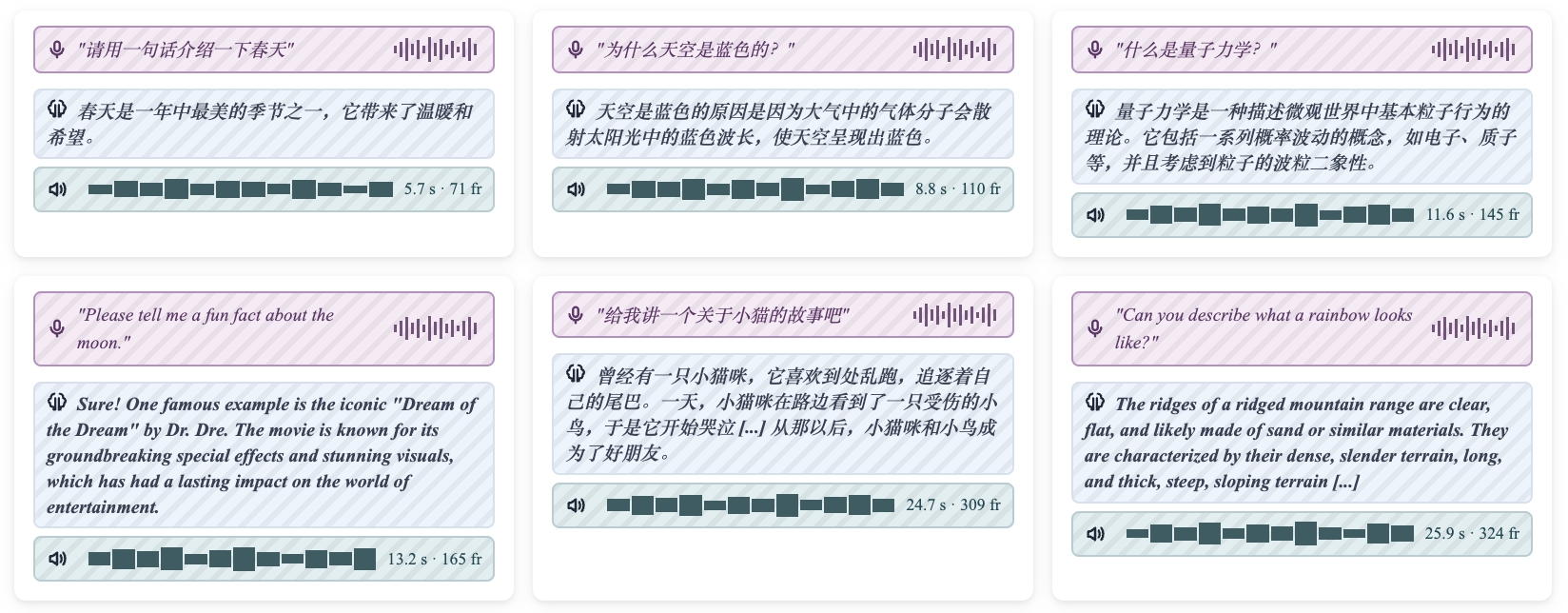}
    \caption{Qualitative A2A examples. The model receives speech input and returns aligned text and speech output, exposing the full speech-in/speech-out loop.}
    \label{fig:qual-a2a}
\end{figure}

\begin{figure}[!ht]
    \centering
    \includegraphics[width=0.99\linewidth]{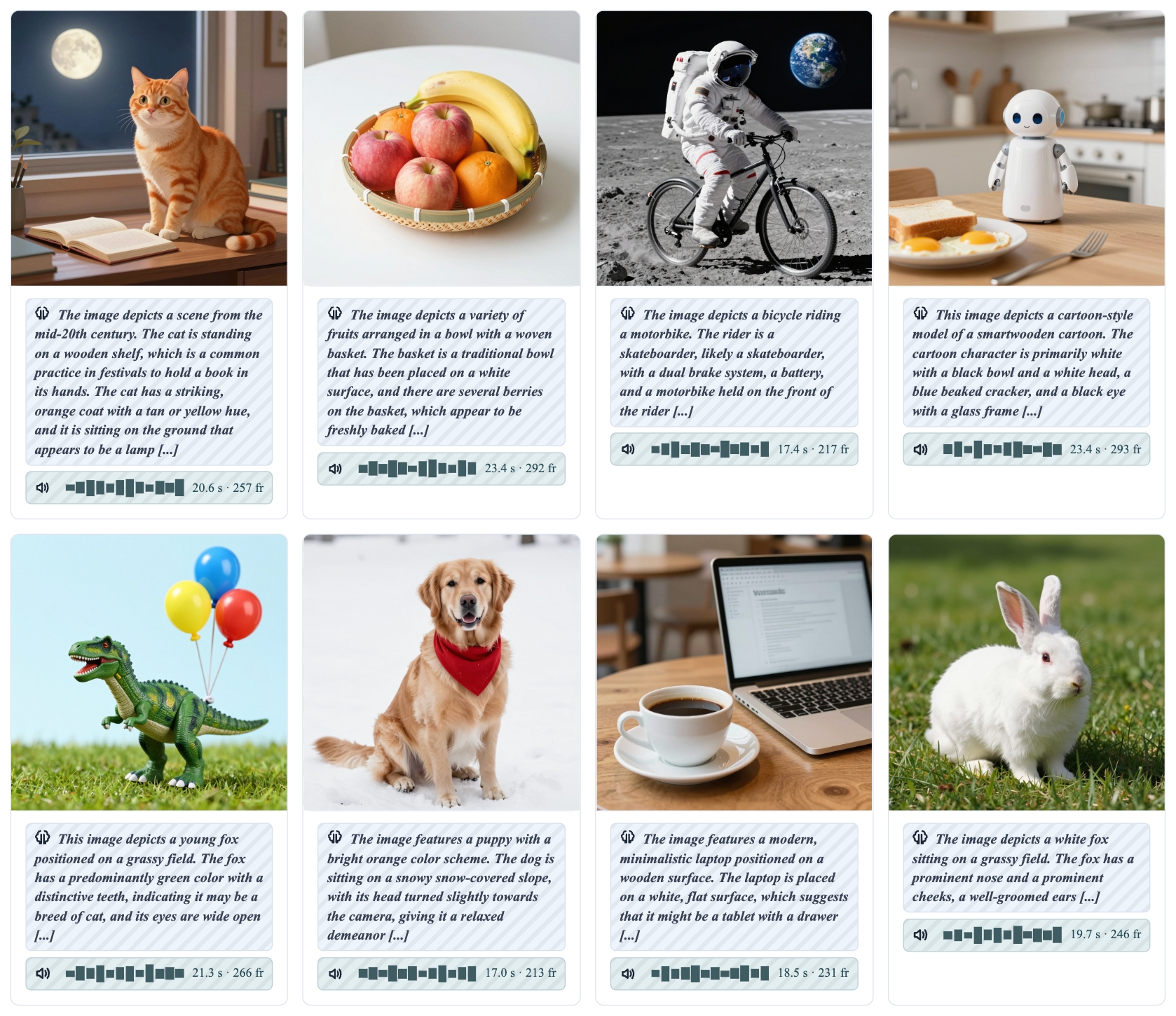}
    \caption{Image-to-audio qualitative examples. Image features are projected into the Thinker, and the resulting answer can be rendered through the Talker as speech.}
    \label{fig:image2audio}
\end{figure}

This appendix shows representative outputs from the three interaction modes supported by \method{}: real-time streaming with barge-in interruption (Figure~\ref{fig:realtime}), audio-to-audio dialogue (Figure~\ref{fig:qual-a2a}), and image-conditioned speech generation (Figure~\ref{fig:image2audio}). The examples are generated by the \dense{} variant, and the HTML demo page bundled with the release includes playable audio for the displayed cases.

Figure~\ref{fig:realtime} shows the real-time streaming and barge-in interaction setting. After the user finishes speaking, the Thinker first performs the semantic-side prefill, the Talker starts producing audio codes, and the Mimi decoder writes the 24\,kHz waveform as new code frames become available. The lower timeline illustrates the barge-in path: when the user speaks again during model playback, the system detects the new speech event, abandons the current generation, and begins a fresh prefill--reply cycle. This is not a claim of human-level full-duplex turn taking; the interrupt detection is still based on a simple VAD threshold rather than semantic understanding of overlap. It is a smaller but practically useful engineering loop: the system can leave the speaking state, accept a new request, and produce the next response without waiting for the previous waveform to finish.

Figure~\ref{fig:qual-a2a} shows audio-to-audio cases where real speech is used as input and the model returns both text and speech. Short assistant-style dialogue is the most stable setting: the Thinker produces a compact semantic answer, and the Talker can render it before audio-code errors accumulate. Chinese explanatory prompts usually remain coherent, while English responses show more variation in pronunciation and rhythm. Longer answers are still possible, but they expose the same weakness as Table~\ref{tab:cross-t2a}: pronunciation drift and small word omissions become easier to trigger as the acoustic path has to sustain a longer sentence.

Figure~\ref{fig:image2audio} illustrates image-conditioned speech generation. The path connects visual encoding, text generation, and speech rendering in a single pipeline: SigLIP2 provides image features, the projector maps them into the Thinker space, and the Talker renders the resulting answer as speech. The examples show that the pipeline can condition speech on image content, but they also expose typical small-model errors: some outputs capture the coarse scene, while others replace the main object or confuse attributes, such as animal categories or vehicle type. These errors are consistent with the 64 image-placeholder budget and the 0.1B base scale, so the examples should be read as evidence that the small omni pipeline runs end-to-end rather than as an upper bound on open-ended image description.
\end{document}